\documentclass{article}
\usepackage{spconf,amsmath,graphicx,booktabs,multirow,verbatim,amssymb}

\title{A Neural Acoustic Echo Canceller Optimized Using An Automatic Speech Recognizer and Large Scale Synthetic Data}
\name{Nathan Howard\sthanks{equal contribution}, Alex Park\footnotemark[1], Turaj Zakizadeh Shabestary, Alexander Gruenstein, Rohit Prabhavalkar}

\address{Google Inc., Mountain View, CA, USA}
\begin{document}
\ninept
\maketitle
\begin{abstract}
We consider the problem of recognizing speech utterances spoken to a device which is generating a known sound waveform; for example, recognizing queries issued to a digital assistant which is generating responses to previous user inputs. Previous work has proposed building acoustic echo cancellation (AEC) models for this task that optimize speech enhancement metrics using both neural network as well as signal processing approaches.

Since our goal is to recognize the input speech, we consider enhancements which improve word error rates (WERs) when the predicted speech signal is passed to an automatic speech recognition (ASR) model. First, we augment the loss function with a term that produces outputs useful to a pre-trained ASR model and show that this augmented loss function improves WER metrics. Second, we demonstrate that augmenting our training dataset of real world examples with a large synthetic dataset improves performance. Crucially, applying SpecAugment style masks to the reference channel during training aids the model in adapting from synthetic to real domains. In experimental evaluations, we find the proposed approaches improve performance, on average, by 57\% over a signal processing baseline and 45\% over the neural AEC model without the proposed changes.
\end{abstract}
\begin{keywords}
Acoustic echo cancellation, deep learning, sequence-to-sequence model, multi-task loss, acoustic simulation
\end{keywords}
\section{Introduction}
\label{sec:intro}
Voice queries have become increasingly common as a way to communicate with smart devices, such as phones and speakers.  In challenging acoustic conditions (background noise, distance from microphone, etc.), interpretation of queries can fail due to poor speech recognition accuracy.  We focus on the problem of acoustic echo cancellation (AEC) -- removing a known source of additive interference.  The term "echo" cancellation is used because the device has access to the original {\bf reference} signal that is the source of interference, but the interference itself is an echoic version of the signal that has propagated through the room before being received at the microphone(s).  In this paper, we will denote the user's speech as the {\bf target} and the mixture of reverberant target and background noise as the {\bf residual}. The received mixture of echoed reference and residual is denoted as the {\bf probe} and the AEC outputs an {\bf erased} signal.

Our end goals are slightly different from typical AEC scenarios because our deployment scenario is echo cancellation in the context of interaction with a smart speaker.  As such, there are two important characteristics of the target signal.  First, we assume that we are attempting to recover a speech signal -- usually a user query.  Second, unlike in telephony or meeting situations, the perceptual fidelity of the recovered signal is not as important as its intelligibility to the ASR system.  With these considerations in mind, we propose a model and training protocol designed to simultaneously perform echo cancellation, dereverberation, and moderate denoising by learning to predict the target signal given the probe and reference signals.

The contributions of this work are as follows:  We propose an autoregressive sequence-to-sequence model for performing acoustic echo cancellation.  We demonstrate the value of optimizing on an ASR encoder loss criterion for producing erased signals which improve intelligibility on ASR systems over purely signal-based metrics.  Finally we implement two methods for improving robustness of the model to distortion between echo and reference: by preparing a mixture of synthetic and quasi-synthetic data for training, and performing  dynamic corruption of the input signals via different configurations of SpecAugment~\cite{Park2019}.

\section{Related Work}
\label{sec:related_work}
In traditional signal processing, linear AEC techniques attempt to estimate the overall system of render-propagation-capture by a time-varying linear filter, usually an adaptive Finite Impulse Response (FIR) filter. Often the filter coefficients are estimated to replicate the echo, in the Minimum Mean Square Error (MMSE) sense, given the reference signal. Then, the filtered version of the reference signal is subtracted from the probe to obtain an estimate of the target signal.

In recent years, there have been numerous proposed approaches to applying neural networks for AEC~\cite{Lei2019, Zhang2018, Fazel2019}.  In most previous work, the criteria for evaluating AEC performance have been signal driven metrics such as signal distortion ratio (SDR), or echo return loss (ERL). Work here often predicts the residual signal by predicting an ideal ratio mask (IRM) that is applied to the probe~\cite{Zhang2018} or gains applied to the output of a linear AEC~\cite{lee2015dnn}. While these metrics are easy to calculate and correlate well to perceptual cancellation quality, our initial experiments indicated that improvements in signal-based metrics often did not translate to proportionally improved WER performance.

Two notable sequence-to-sequence speech prediction models that have been proposed recently are Parrotron~\cite{biadsy2019parrotron} and Textual Echo Cancellation~\cite{ding2020textual}. The authors in~\cite{biadsy2019parrotron} use an ASR encoder and a text-to-speech (TTS) decoder to perform speech transformation.  In order to optimize for intelligibility, the Parrotron model is trained to simultaneously minimize an ASR decoding loss as well as a spectral decoding loss on the same encoded representation.  A drawback of this approach is that the transcription of the source signal is required in order to compute the ASR related loss.  In~\cite{ding2020textual}, the authors assume that the echoed reference is generated by text-to-speech (TTS) and use a Parrotron-style network to remove the echoed reference using only the textual source of the reference signal.  The model in that paper uses only spectral loss for training.

\section{Model Architecture}
The proposed neural AEC model uses an encoder-decoder structure to reconstruct spectral frames of the erased signal by casting the problem as a sequence-to-sequence task.  As in~\cite{biadsy2019parrotron} and~\cite{ding2020textual}, we use frame level features (80-dimensional log-mel spectral vectors) for both source and target sequences.  The source sequence features are computed from the probe and reference signals, and the target sequence features are computed from the clean target signal.  Although all three signals should be synchronous, in this system we align probes and references using their cross correlation and enforce that source and target sequences have matching lengths.

The model is comprised of a speech encoder followed by a spectral decoder, which are described in the following sections.

\subsection{Encoder}
The speech encoder is similar to the encoder described in~\cite{He2018-RNN-t}, which takes a sequence of speech features as input and produces a high dimensional hidden representation sequence.  We compute feature frames for each of the probe and reference signals, then stack each frame depthwise to create an input tensor that has shape $[B, T, 80, 2]$, where $B$ is the batch size and $T$ is the number of frames.  For the encoder used in this work, we used 3 unidirectional LSTM layers, each with 512 hidden dimensions, and no temporal downsampling, so the number of hidden representation has the same number of frames as the input.

\subsection{Decoder}
We use an autoregressive spectral decoder to predict a sequence of spectral frames from the encoded sequence.  The decoder is based on the decoder component described in~\cite{Shen2017-Tacotron}, which is designed to produce spectrogram frames.  For the decoder used in this work, we made two small changes.  First, because the context needed to transform input to output is local to the frame being processed, we omitted the attention layer.  Also, since we constrain the output to be the same number of frames as the input, we also omit the end-of-sequence prediction component of the decoder -- the decoder stops producing input when the input frames are exhausted.

The spectral decoder consists of a single 512 dimension LSTM layer followed by an 80 dimension projection layer that feeds its output to a pre-net and a post-net.  The pre-net is a feed-forward network that serves to gate the influence of the previous time-step's output compared to the source. The post-net is a stack of five convolutional layers that act on the predicted spectral frames to produce a residual correction factor that is added to create the final prediction.  Each non-final convolutional layer applies a 1-dimensional convolution in time, with 512 filters of size 5, followed by batch normalization and tanh activation.  The output of the decoder is a sequence of 80-dimensional log-mel spectral frames which can be inverted back to a time  domain waveform via Griffin-Lim~\cite{Griffin1984} or by using a neural vocoder~\cite{oord2017parallel}.  For this paper, we used Griffin-Lim to produce waveform inputs when needed (e.g. for performing recognition evaluations on AEC output).

\begin{figure}[tb]
\begin{minipage}[b]{1.0\linewidth}
  \centering
  \centerline{\includegraphics[width=8.5cm]{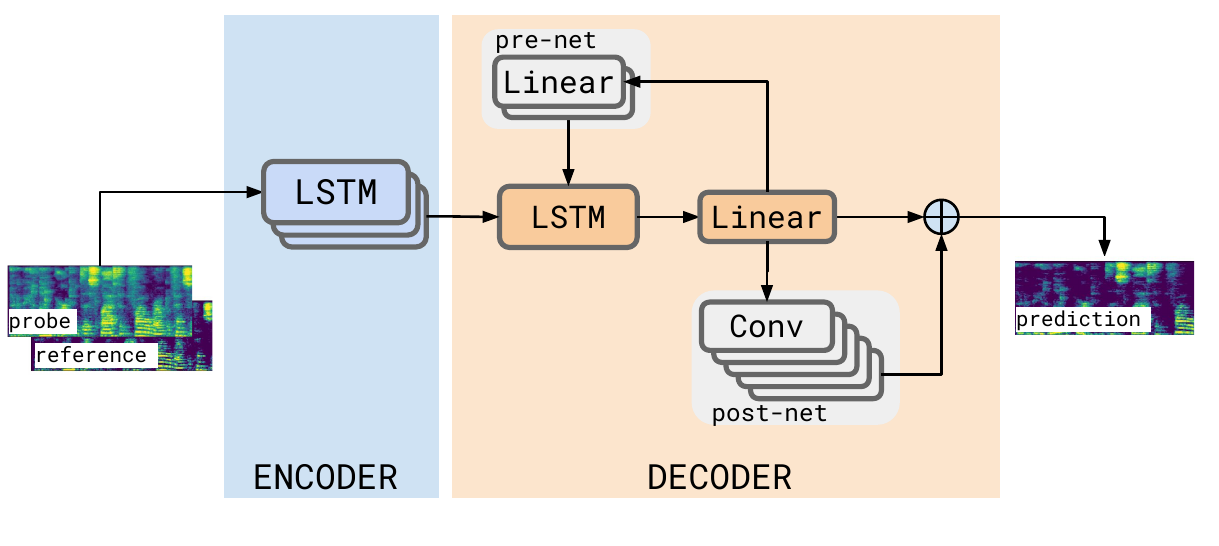}}
\end{minipage}
\caption{Model block diagram for inference pathway.  The encoder, structured as a typical speech encoder, takes in frame level features and produces a latent representation that is decoded by a spectral decoder to produce frame level features.}
\label{fig:model}
\end{figure}

\subsection{Loss Function}
Our initial experiments used purely spectral loss when training the network.  This loss is computed by summing the mean $L1$ and mean $L2$ (or MSE) distance between the target spectral features and the output of the decoder, both before and after applying the post-net correction.

Since the motivation for our work is improving speech recognition accuracy on the erased signal predicted by our AEC, we also explored changing the loss function to bias the AEC towards producing outputs useful to an ASR as input. In the ideal case, both the AEC output and target signal would produce the same latent representation when run through the ASR model's encoder. Observing this, we integrated an ASR loss which runs the predicted and target features through an ASR encoder, pre-trained on clean audio, and computes the MSE between the respective latent representations.  The final loss term is then
\[{\tt loss}\; =\; {\tt loss}_{spectral} + \lambda\; {\tt loss}_{ASR}\]
where $\lambda$ is a hyperparameter.  Figure~\ref{fig:res} illustrates how the losses are computed and combined.  Unlike the auxiliary decoder loss used by Parrotron, the ASR encoder loss compares latent representations of the predicted and target signals without needing the underlying transcription, which removes the need for labeled training data.

\subsection{Training}
\label{ssec:training}
All of our models are trained using Lingvo~\cite{DBLP:journals/corr/abs-1902-08295}, which is built on top of TensorFlow~\cite{DBLP:journals/corr/AbadiABBCCCDDDG16}. The AEC models were trained with a batch size of 128 using the ADAM optimizer~\cite{kingma2017adam} and scheduled sampling~\cite{bengio2015scheduled} on a randomly selected half of the autoregressive decoder input.

The ASR model used in the loss function is a ContextNet~\cite{han2020contextnet} CNN-RNN transducer, trained on LibriSpeech to 3.8 WER on {\tt test-clean}. The whole model has 31 million parameters and the encoder contains 23 stacked convolution blocks.  Importantly, the AEC model fails to converge when the ASR loss is included at the start of training. We resolve this by making $\lambda$ dependent on the current training step and linearly ramping $\lambda$ up from zero to its final value, 0.01, over the first 20k training steps.

Because SpecAugment has been shown to be a useful method of data augmentation for improving WER performance~\cite{Park2019, biadsy2019parrotron}, we also experimented with applying SpecAugment to  AEC model inputs during training. When using SpecAugment, we masked up to 27 of 80 frequency bins divided between 2 frequency masks and up to 5\% of frames split between 10 time masks. Models using SpecAugment trained for 200k steps and models without for 90k steps.
 
\section{Data Preparation}
A key challenge in building a neural network based AEC is data collection. In real world recordings, the echoed reference component of the probe can be distorted by non-linearities in the loudspeaker's reproduction of the signal~\cite{Enzner2014}. These distortions can vary at different volumes, temperatures and between different loudspeakers.  A common practice is to apply a functional non-linearity to mimic loudspeaker distortion as in~\cite{chen2020nonlinear}.  Of course, the highest fidelity way to capture these effects in training data is to record echoed reference outputs in real rooms, but this has the considerable downside of being expensive and time consuming.  We used a multi-pronged approach to creating diverse AEC training data - by processing with a room simulator, by combining re-recorded real world data with a room simulator, and by dynamically augmenting the data during training using SpecAugment~\cite{Park2019}. 
\vspace{-.1in}
\subsection{Source Data}
For training and evaluation of the AEC techniques compared in this paper, we drew from two sources of speech data: parts of the LibriSpeech corpus were used as both targets and references, and an internal set of TacoTron-generated~\cite{Shen2017-Tacotron} TTS utterances were used as references.  For simulating room environments, we used the room simulator described in~\cite{Kim2017b}.  Separately to the echoed reference, background noise was added as described in~\cite{Kim2017b}, with noise sources drawn from a set of daily life and cafe noise recordings.  

\subsection{Training Data}
\subsubsection{Synthetic Echo}
\label{ssec:synthetic_reference}
In this setup, the return path of the echoed reference was wholly simulated.  The target and reference signals were randomly selected from the {\tt train-clean} portion of the LibriSpeech corpus.  For each synthetically noisified utterance, a room configuration was sampled from one of 100,000 possibilities, and the simulated probe, echoed reference, and residuals were computed via simulation.  The room configurations were constrained to replicate the geometry of a smart speaker; with the loudspeaker set up as a noise source in a fixed position relative to the microphones.  The target source was randomly positioned away from the microphones, with elevation angle restricted to the interval $[45^{\circ} , 135^{\circ} ]$ and distance varying between 0.25 and 8 meters, with a mean of 2.5 meters.  For this simulated condition, the target-to-noise-ratio and target-to-echoed-reference-ratio were randomly chosen in the ranges of (0, 20) and (-20, 0) dB, respectively.  In total, there were approximately 153k training utterances produced using the synthetic echo setup.

\subsubsection{Re-recorded Echo}
\label{ssec:rerecorded_reference}
In order to account for real loudspeaker-induced distortions and differences between synthetic room impulse responses and real-world room return path effects, we also created a set of re-recorded echo utterances.  Drawing from the TTS utterances, we collected re-recorded versions of these utterances as echoed reference signals by playing them out of smart speakers in various conference room environments and recording the resulting output on the smart speaker microphones.  7592 training pairs and 1546 test pairs of (reference, echoed reference) signals were collected in this manner.

The re-recordings were then combined with target signals drawn, without re-recording, from the {\tt train-clean} portion of LibriSpeech using the same room configurations as in Section~\ref{ssec:synthetic_reference}.  The re-recorded echo signal was used directly as the echoed reference, without propagating through the room simulation.  Otherwise, the echoed reference, background noise source and simulated path of the target signal were mixed together with the same distribution of SNRs as in Section~\ref{ssec:synthetic_reference}.  Approximately 34k training utterances were produced using this re-recorded setup.

\subsection{Evaluation data}
The test sets for evaluation were constructed as described in Section~\ref{ssec:rerecorded_reference}, but with the test pairs of re-recorded reference signals, target signals drawn from LibriSpeech {\tt test-clean}, and target-to-echoed-reference-ratio levels held fixed at 0dB, -5 dB, and -10 dB to create three test set variants of escalating difficulty. The room impulse responses and background noise samples used for the test sets were all unseen during training. 

\begin{figure}[t]

\begin{minipage}[b]{1.0\linewidth}
  \centering
  \centerline{\includegraphics[width=8.5cm]{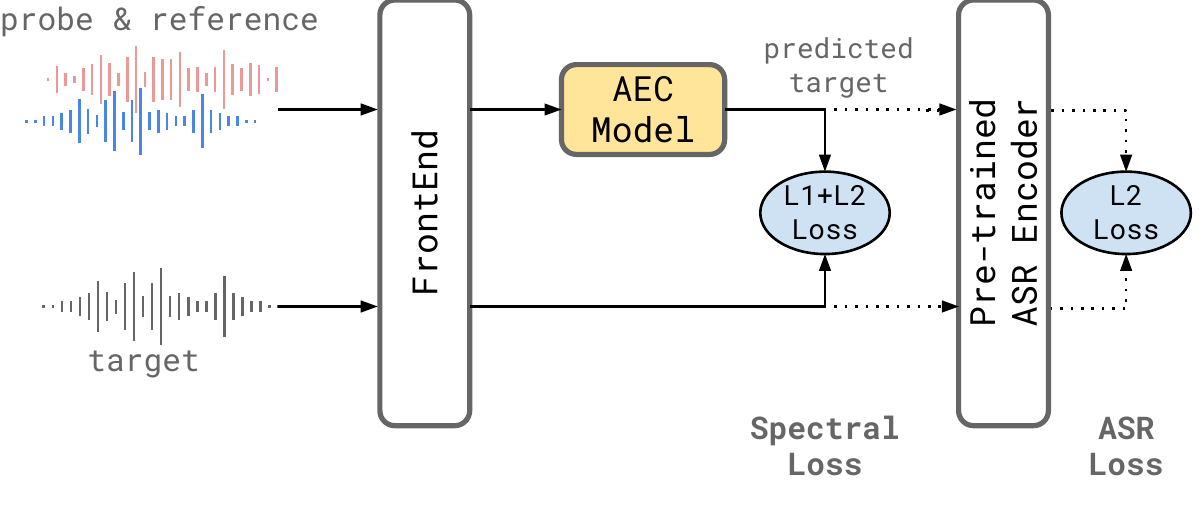}}
\end{minipage}
\caption{Two types of losses are are used to optimize the AEC model.  Spectral loss is computed between the predicted output and ground truth target features.  ASR loss is computed between the encoded representations of the predicted features after passing through a pre-trained ASR encoder.  ASR encoder weights are kept fixed while training the AEC model.}
\label{fig:res}
\end{figure}
\vspace{-.1in}
\section{EXPERIMENTS}
\label{sec:experiments}
Unless otherwise specified, speech recognition results were obtained using the ContextNet ASR model described in Section~\ref{ssec:training}.

\subsection{Data Augmentation Effects}
\label{ssec:data_effects}
We varied the inputs used for training to gain insight into the effects of augmenting datasets and inputs.  These results are shown in Table~\ref{tbl:domain_adaptation}.  When controlling for dataset and loss function, we found that applying SpecAugment to the reference signal alone resulted in the most consistent WER reductions.  This was the SpecAugment configuration used in our final model during training.  Our interpretation of this outcome is that SpecAugment introduces a challenging form of mismatch between the reference and its echo for which the model must compensate and that this mismatch is different from and complementary to the diversity of echoic effects presented by the synthetic/re-recorded data alone.

By looking at matched SpecAugment configurations in Table~\ref{tbl:domain_adaptation}, we observe the benefit of including synthetic {\em and} re-recorded data in training.  Although the re-recorded training data is closest to test set conditions, there are still significant gains from adding training set diversity.  When not applying SpecAugment, there was a relative WER reductions of 25.5\% (averaged across SNR levels) when using the larger combined dataset compared to the re-recorded data alone.

\begin{table}[t]
\centering
\begin{tabular}{@{}ll|lll@{}}
\toprule
Training Dataset                        & SpecAugment       & 0dB   & -5dB  & -10dB \\ \midrule
\multirow{4}{*}{Synthetic Reference}    & None              & 59.14 & 70.33 & 80.75 \\
                                        & Both Inputs       & 47.68 & 59.26 & 71.14 \\
                                        & Probe Only        & 53.94 & 66.15 & 78.01 \\
                                        & Reference Only    & 31.47 & 43.23 & 57.50 \\ \midrule
Re-recorded Reference                      & None              & 19.66 & 25.35 & 34.28 \\ \midrule
\multirow{4}{*}{Synthetic + Re-recorded}  & None              & 14.19 & 18.71 & 26.54 \\
                                        & Both Inputs       & 13.13 & 15.97 & 21.98 \\
                                        & Probe Only        & 12.79 & 15.83 & 22.15 \\
                                        & Reference Only    & 12.51 & 15.54 & 22.30 \\ \bottomrule
\end{tabular}
\caption{Input data effects. WERs of models trained with different SpecAugment configurations and dataset partitions.  All models here were trained with {\bf spectral loss only}.}
\label{tbl:domain_adaptation}
\end{table}

\subsection{ASR Loss Robustness}
\label{ssec:asr_loss_robustness}
One concern with optimizing the AEC model using a pretrained ASR encoder is that the AEC will overfit to the idiosyncrasies of that specific ASR encoder and produce outputs that are mismatched when presented to other ASR models. To measure this effect, we trained two AEC models, one with and one without ${\tt loss}_{ASR}$, and evaluated the outputs on three different ASR models, of which only {\bf one} was used for calculating ${\tt loss}_{ASR}$ during training.

The three pre-trained ASR models evaluated are:  a CNN-based global context model~\cite{han2020contextnet},  a bidirectional LSTM-based listen-attend-spell model~\cite{chan2015listen_a}, and a streaming LSTM-based RNN-T model~\cite{sainath2020streaming}.  The first two models were trained on the train partition of LibriSpeech, and the last model was trained on a large corpus of far field and near field non-LibriSpeech utterances. 

\begin{figure}[tb]

\begin{minipage}[b]{1.0\linewidth}
  \centering
  \vspace{-.2in}
  \centerline{\includegraphics[width=8.5cm]{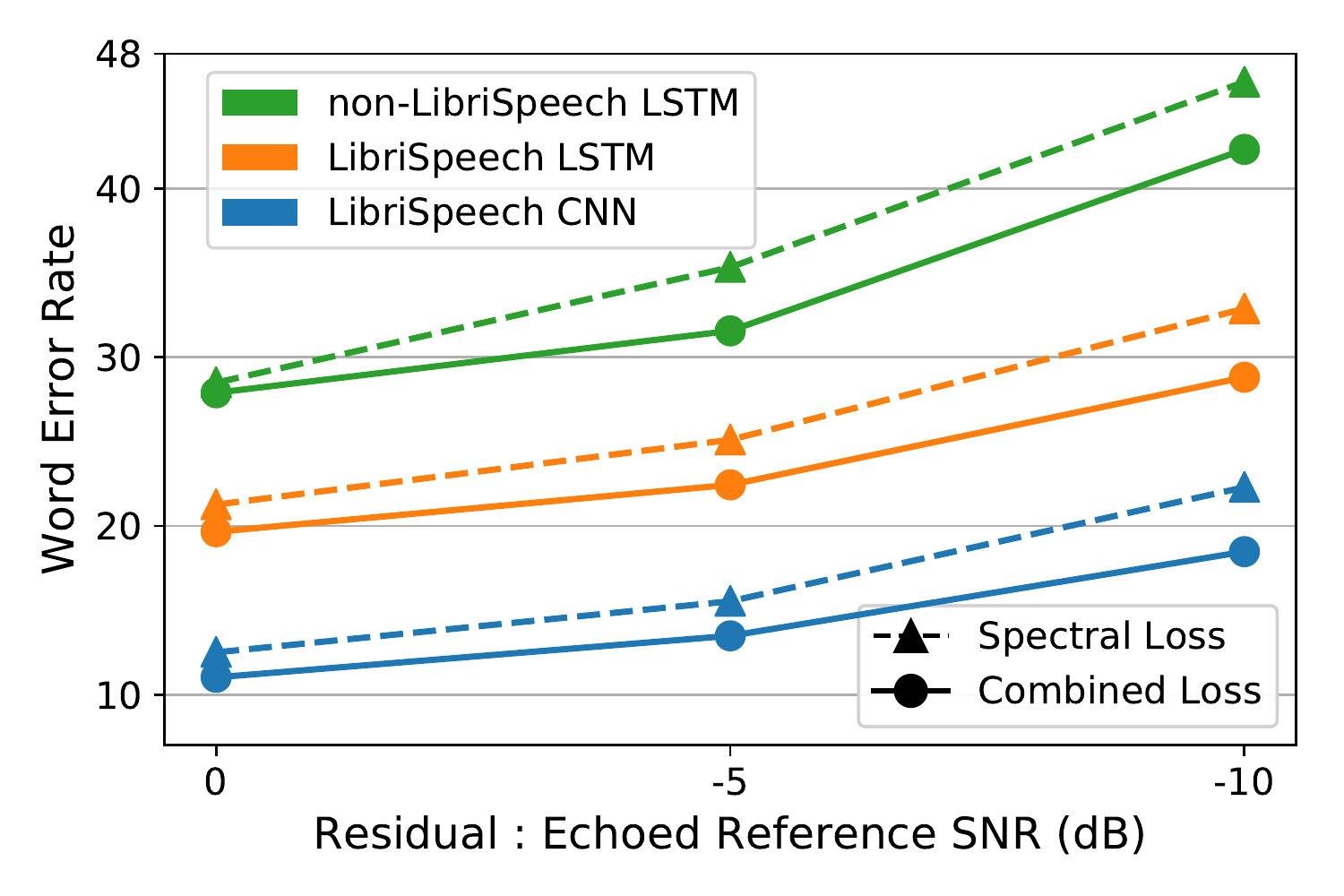}}
\end{minipage}
\caption{ASR Loss Robustness.  Word Error rates for different ASR models on AEC model outputs, comparing outcomes when training with and without ${\tt loss}_{ASR}$.  Only the LibriSpeech CNN model (blue line) was used for computing ${\tt loss}_{ASR}$ during training.}
\label{fig:asr_comp_chart}
\end{figure}

Figure~\ref{fig:asr_comp_chart} shows the WER of each of the ASR models on the outputs of the AEC models.  Though only the in-domain CNN speech encoder was used to calculate ${\tt loss}_{ASR}$ for the AEC model incorporating that loss, we observe consistent improvements for the other two ASR models across all SNRs as well.  This improvement holds despite significant differences in model structure, training data, and frontend configuration.  As expected, we observe the largest improvements for the matched ASR encoder (CNN), followed by the in-domain LSTM recognizer, and smaller, but still significant gains for the out-of-domain model.

\subsection{Final System}
We combined all of the proposed modifications to the model and evaluated WER results in Table~\ref{fig:res}.  For comparison purposes, we contrast against two other AEC techniques.  The first is a linear AEC system that performs adaptive filtering on STFT subbands, similar to~\cite{avargel2006}, but using longer STFT frames and within-band only filter taps.  We also implemented and trained a mask-based neural network AEC model as described in~\cite{Zhang2018}.  That model is trained to predict an ideal ratio mask (IRM) that is then used to mask the spectral magnitude of the probe, which is then inverted back to the time domain.  During training, the IRM target is computed using the residual and echoed reference.  When training this model, we used both synthetic and re-recorded data, but did not apply SpecAugment.

As expected, all AEC techniques significantly improve recognition accuracy compared to evaluating on the probe signal alone.  Moreover, both neural models improve over the STFT-based AEC at higher SNRs (0 dB and -5 dB), but the IRM-based model degrades much more sharply than the STFT-based AEC as SNR decreases.  Our neural model, when including all proposed improvements, achieves significant improvements compared to both alternatives at all three SNR levels.  In addition to the analysis in Sections~\ref{ssec:data_effects} and~\ref{ssec:asr_loss_robustness}, we show ablation results from successively removing each of the proposed improvements from the final system.  

Interestingly, our final model produces outputs that yield better WER in the 0 dB case than running recognition on the residual signal directly, which has no echoed reference. This is presumably because the model was trained with non-reverberant, noise-free utterances as its training targets and therefore learned to predict de-reverberated and de-noised features rather than just the residual.

\begin{table}[tb]
\centering
\begin{tabular}{@{}lllll@{}}
\toprule
                                                        & 0 dB  & -5 dB & -10 dB            \\ \midrule
\multicolumn{1}{l|}{Target*}                             & \multicolumn{3}{c}{--- 3.81 ---}  \\
\multicolumn{1}{l|}{Residual*}                           & 12.18 & 11.14 & 12.30             \\
\multicolumn{1}{l|}{Probe}                              & 75.86 & 82.20 & 85.78             \\
\multicolumn{1}{l|}{STFT-based AEC}                     & 31.37 & 32.55 & 36.91             \\
\multicolumn{1}{l|}{IRM AEC~\cite{Zhang2018}}           & 23.01 & 30.75 & 41.85             \\
\multicolumn{1}{l|}{Neural AEC\:(ours)}                 & \textbf{11.03} & \textbf{13.49} & \textbf{18.48}  \\
\multicolumn{1}{l|}{\:\:\:-AsrLoss}              & 12.51 & 15.54 & 22.30             \\
\multicolumn{1}{l|}{\:\:\:\:\:\:-SpecAugment}          & 14.19 & 18.71 & 26.54             \\
\multicolumn{1}{l|}{\:\:\:\:\:\:\:\:\:-Synthetic Dataset}    & 19.66 & 25.35 & 34.28             \\ \bottomrule
\end{tabular}
\caption{System comparison.  WER calculated on different SNR test subsets using various AEC models and available signals.  Target and residual are oracle signals not available to the model at inference.}
\label{tbl:wer}
\end{table}

\section{CONCLUSION}
\label{sec:conclusion}
We proposed an autoregressive neural network model to perform AEC in situations with double-talk and background noise.  The model was trained using a dataset augmented with synthetic examples with SpecAugment masks applied to increase robustness to mismatch between the reference and the echoed reference. To adapt the model towards being an input to an ASR system, the loss function was extended with a pretrained ASR encoder. When compared to a purely signal processing-based AEC technique and a mask-based neural AEC model, our proposed approach improved speech recognition accuracy across several noise levels.

\vspace{.1in}
\noindent
{\bf Acknowledgements:} Thanks to James Walker and Bharath Ranganatha Mankalale for aid in data collection and room simulation.

\vfill\pagebreak

\bibliographystyle{IEEEbibAlt}
\bibliography{refs}

\end{document}